\documentclass[reprint,
unsortedaddress,
twocolumn,
amsmath,
amssymb,
aps,
prl,
]{revtex4-1}
\usepackage{braket}
\usepackage{graphicx}
\usepackage{lipsum}
\usepackage{dcolumn}
\usepackage{bm,color}
\usepackage[normalem]{ulem}

\usepackage{amsfonts}

\usepackage{epigraph}

\usepackage{dsfont}
\usepackage{etoolbox}
\makeatletter
\newlength\epitextskip
\pretocmd{\@epitext}{\em}{}{}
\apptocmd{\@epitext}{\em}{}{}
\patchcmd{\epigraph}{\@epitext{#1}\\}{\@epitext{#1}\\[\epitextskip]}{}{}
\makeatother

\begin{document}

\title{The Quantum Foucault Modes}

\author{Samuel Alperin}

\affiliation{\vspace{1.25mm} \mbox{Los Alamos National Laboratory, Los Alamos, New Mexico 87545, USA}}

\begin{abstract}

The driven quantum harmonic oscillator is fundamental to a number of important physical systems. Here, we consider the quantum harmonic oscillator under non-Hermitian, PT-symmetric driving, showing that the resulting set of Wigner-space trajectories of an initial coherent state is identical to the set of real-space trajectories of the classical Foucault pendulum. Remarkably, in the case mapped from the trivial 1D pendulum, the corresponding quantum dynamics are those of an oscillator with periodically evolving momentum but fixed position, a novel type of dynamics which are forbidden in classical systems. 

\end{abstract}

\maketitle

There are few models in physics as broadly fundamental as the harmonic oscillator,
from simple systems of classical springs to quantum field theories. In both classical and quantum settings, the simplest and perhaps most important generalization of the simple harmonic oscillator is the addition of a driving force. For the quantum case, the dynamics of an initial coherent state evolves in the Wigner space exactly as the corresponding classical system evolves in classical phase space: for suitably off-resonant driving, the dynamics follow an elliptical trajectory, while for near-resonant driving the system becomes unstable. \cite{kovsata2022fixing}.
Usually, one would consider this system fully characterized, and would move on to Hamiltonians with additional terms.

However, two decades ago Bender and Boetcher showed that non-Hermitian Hamiltonians, such as those of the form $H= \frac{\hat{p}^2}{2 m}+\frac{m}{2}\omega_{0}^2 \hat{x}^2 - i\hat{x}$, can be physical, and have strictly real spectra \cite{bender1998real}. Since that realization it has been found that many simple Hamiltonians, transformed under analytic continuation into the complex plane, have strikingly different, often exotic behaviours such as topological skin effects and unidirectional invisibility \cite{lin2011unidirectional,xiao2017observation,bandres2018topological,harari2018topological,xia2021nonlinear}. 
At the same time there has been a dramatic increase in the ability of various experimental systems to physically realize nearly arbitrary complex potentials \cite{miri2019exceptional,el2018non}, including those driven by explicitly time-dependent, complex-frequency waves \cite{kim2023loss}.

In this work, we consider the dynamics of the quantum harmonic oscillator under a particular class of complex driving potentials, showing that the addition of an imaginary component to the driving potential such that $F(t)=F_0\big(\cos(\omega t)\pm i\sin(\omega t)\big)$ leads to a class of quantum dynamical Wigner space trajectories which we show correspond exactly to the real space trajectories of the classical Foucault pendulum. From this duality, we find several particularly important classes of dynamics, demonstrating the explicitly non-Hermitian stability of the quantum harmonic oscillator driven at resonance, finding an infinite class of circular orbits with frequencies that are not fixed by that of the harmonic potential, and a class of  quantum trajectory which has no classical analogue: an oscillator with fixed position but periodically evolving momentum. We discuss the physical meaning of this novel class of quantum state.

We begin with the dynamics of the general driven quantum Harmonic oscillator, which are described by the Hamiltonian 
\begin{equation}
    H = \frac{\hat{p}^2}{2 m}+\frac{m}{2}\omega_{0}^2 \hat{x}^2 - F(t)\hat{x}
\end{equation}
with mass $m$, resonant frequency $\omega_0$, and periodic function $F(t)$ with frequency $\omega$. $\hat{x}$ and $\hat{p}$ denote the position and momentum operators, respectively. Such a Hamiltonian can be physically realized in a number of systems, including both the laser-driven Fabry-Perot cavity \cite{scully1997quantum,loudon2000quantum,barnett2002methods} and the microwave-driven LC circuit \cite{blais2021circuit}, systems which represent the basis of cavity and circuit QED, repsectively, and which are thus fundamental to quantum information and metrology. Defining the annihilation and creation operators as $\hat{a}=\sqrt{\omega_0m/2\hbar}(\hat{x}+i\hat{p}/\omega_0m)$ and $\hat{a}^\dagger=\sqrt{\omega_0m/2\hbar}(\hat{x}-i\hat{p}/\omega_0m)$ , it is useful to rewrite the Hamiltonian as
\begin{equation}
    H = \hbar\omega_0 (\hat{a}^\dagger \hat{a}+1/2)-F(t)(\hat{a}^\dagger+\hat{a})
    \label{ham1}
\end{equation}
 where we have rescaled $F(t)\rightarrow\hbar F(t)/(2\sqrt{2\hbar m\omega_0})$. This Hamiltonian can be solved exactly in more than one way, but for a flexible treatment of the time-dependent driving, we use the Lie-algebraic decoupling methods developed by Wei and Norman \cite{wei1963lie,wei1964global}.
 Eq. \ref{ham1} is generated by a finite-dimensional Lie algebra with elements $\{\hat{a}^{\dagger}\hat{a},\hat{a}^{\dagger},\hat{a}, \mathds{1} \}$ and commutators $\{[\hat{a},\hat{a}^{\dagger}] =\mathds{1},[\hat{a}^{\dagger}\hat{a},\hat{a}]=- \hat{a},[\hat{a}^{\dagger}\hat{a},\hat{a}^{\dagger}]=\hat{a}^{\dagger} \}$. A number of works have detailed the use of these techniques to analyze Hamiltonians of this form \cite{uskov2006geometric,kinsler1991quantum,qvarfort2025solving}, which we follow here. Defining the time-evolution operator 
\begin{equation}
    \hat{U}(t)=\mathcal{T}\mathrm{exp}\left[\frac{-i}{\hbar}\int^t_0 \hat{H}(t)dt^{'}\right]
\end{equation} where $\mathcal{T}$ is the time-ordering operator, one can make use of the the Wei-Norman ansatz 
\begin{equation}\label{wei}
     \hat{U}(t)=e^{-if_0(t)\mathds{1}}e^{-if_1(t)\hat{a}^{\dagger}\hat{a}}e^{-if_2(t)\hat{a}^{\dagger}}e^{-if_3(t)\hat{a}}.
\end{equation}
Following the derivation of Qvafort and Pikovski \cite{qvarfort2025solving}, we differentiate $\hat{U}(t)$ with respect to time and multiplying on the right by $\hat{U}^{-1}(t)$, and note that
\begin{equation} \label{eq1}
\begin{split}
\frac{d}{dt}\hat{U}(t)\hat{U}^{-1}(t) &= -i\hat{H}(t)  \\
 & =-i\partial_tf_0-i\partial_tf_1\hat{a}^{\dagger}\hat{a}-i\partial_tf_2e^{-if_0}\hat{a}^{\dagger}\\& \textcolor{white}{=}-if_3(e^{if_0}\hat{a}+if_3)
\end{split}
\end{equation}
This defines a set of differential equations constraining the unknown time-dependent functions $f_n$, which go as

\begin{equation} \label{deq}
\begin{split}
\partial_tf_0 & = -if_2\partial_tf_3
\\\partial_tf_1 &= 1  
\\\partial_tf_2 & = F(t)e^{f_1 t}
\\ \partial_tf_3& =F^*(t)e^{-f_1 t}.
\end{split}
\end{equation}
Choosing the condition $F(0)=0$, Eqs. \ref{deq} can be solved explicitly, so that 

\begin{figure}[t!]
\centering
\includegraphics[width=\columnwidth]{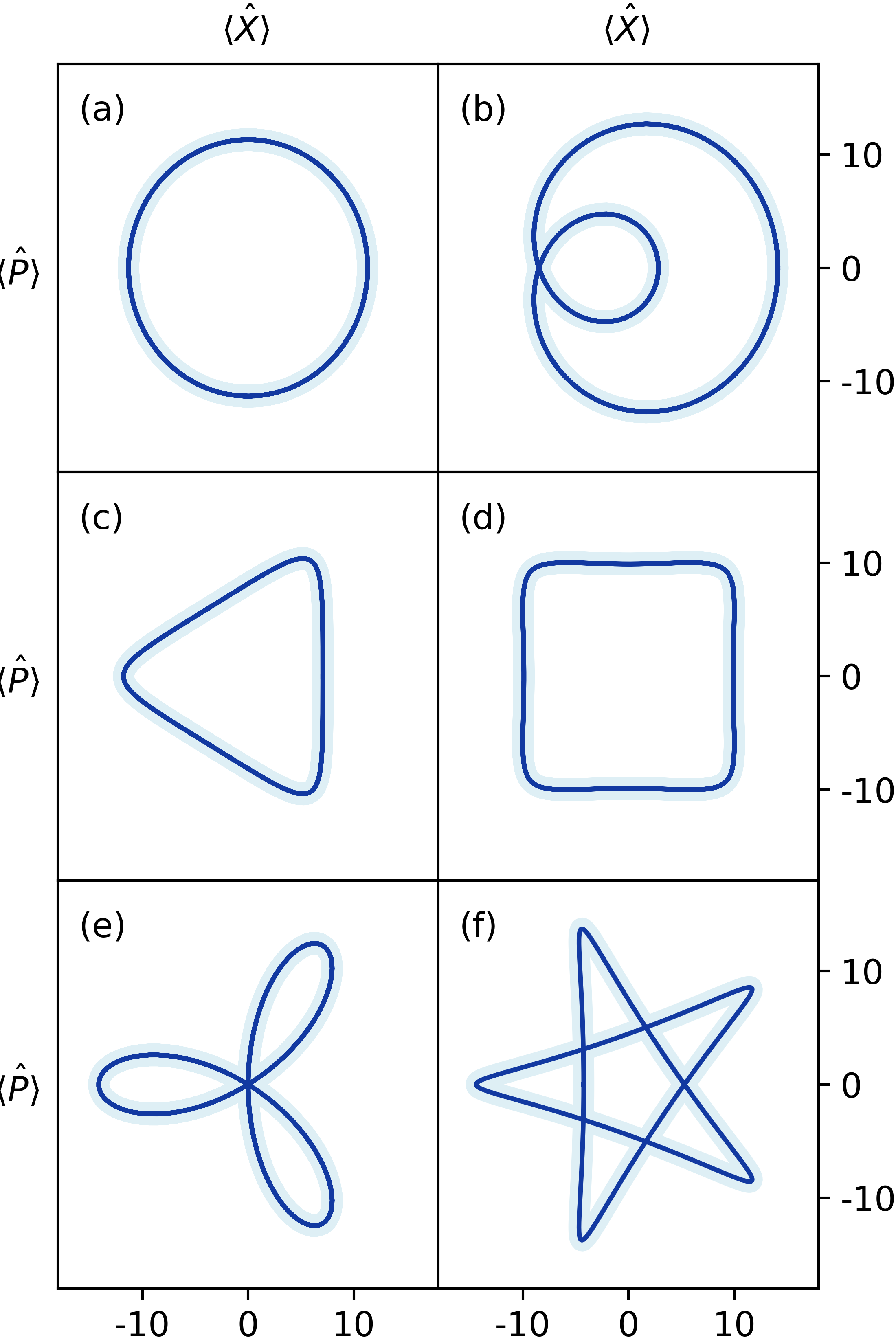}
\caption{Examples of the exact quantum trajectory of a coherent state evolving under PT-symmetric driving. The solid lines represent the paths of the expectation values of the quadratures $\hat{x}$ and $\hat{p}$, with variance represented by semi-transparent regions. Some special cases include: (a): the frequency doubled circular orbits $\big(\omega=-2\omega; F_0=n_0=8\big)$, (b): the cardiods $\big(\omega=-2\omega; F_0=6; n_0=10\big)$, (c) and (d): the rounded polygons $\big(\omega=2\omega; F_0=n_0=5\big)$ and $\big(\omega=3\omega; F_0=4.5; n_0=7\big)$, (e): the rose curves $\big(\omega=2\omega; F_0=15;n_0=0\big)$, and the pentagram $\big(\omega=2\omega/3; F_0=-3;n_0=0\big)$
}\label{fig1}
\end{figure}
\begin{equation} \label{inteq}
\begin{split}
f_1(t)&=t\\
f_2(t)&=f_3^{*}(t)  = \int_0^tF(t^{'})e^{-it^{'}}
\\f_0(t)& =-i\int_0^tF^{*}(t^{'})e^{-it^{'}}dt^{''}\int_0^{t^{'}}F(t^{''})e^{it^{''}}dt^{''}.
\end{split}
\end{equation}
As a result, the time evolution operator given by Eq. \ref{wei} is fully defined by the choice of driving function $F(t)$.

\begin{center}
\begin{figure*}[t!]
\includegraphics[width=\textwidth]{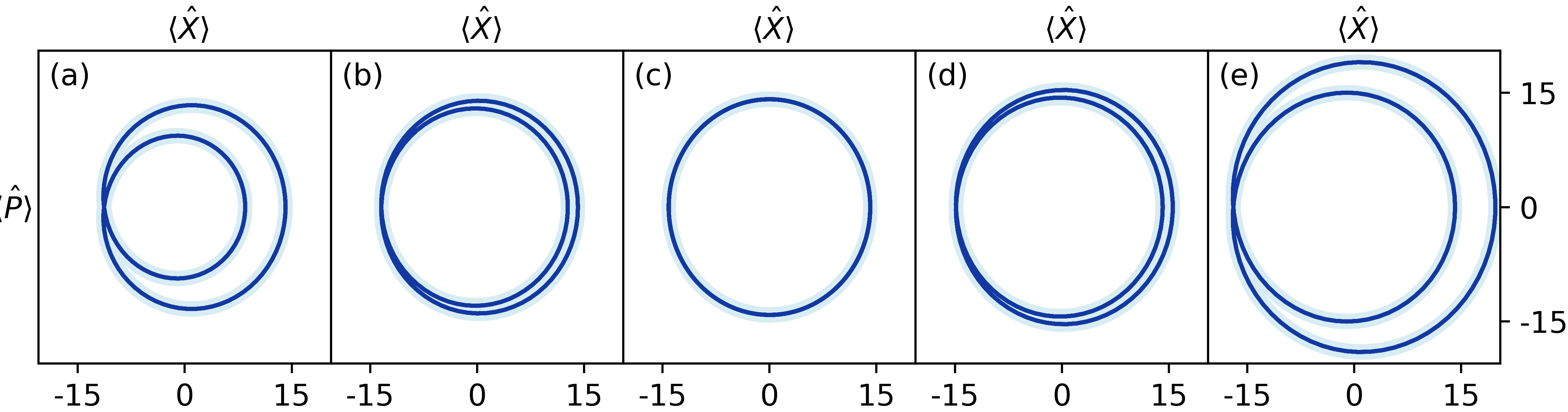}
\caption{Quantum trajectories demonstrating cardiod solutions. As driving strength $F_0$ is increased (from left) with all other parameters fixed, it can be seen that at a particular driving strength, the two lobes of the cardiod overlap to form a single circular path, which is traversed twice by the quantum state within a single natural period of the system. For forcing at frequency $-\omega$ and any fixed set of system parameters ${n_0,\omega_0}$, there is exactly one driving strength $F_0$ for which the Wigner space trajectory is exactly circular, but with frequency $\omega$. In this figure, $\omega=-2 \omega_0$ and $n_0=10$, and from left, $F_0\in\{8,9.5,10,10.5,12 \}$.}\label{freq_doub}
\end{figure*}
\end{center}

From this point, we diverge from what has been studied previously by allowing this drive to be of more general PT-symmetric form, setting $F(t)= F_0\big(\cos(\omega t)\pm i \sin(\omega t)\big)=F_0\exp(\pm i\omega t)$ \footnote{The fully general PT-symmetric drive would be of the form $F(t)= F_0\big(\alpha \cos(\omega t)+ i \beta\sin(\omega t)\big)$. Such a drive yields a broader class of closed-form solutions to the time dependent expectation values $\langle\hat{x}(t)\rangle$ and $\langle\hat{x}(t)\rangle$, but for clarity those solutions are not considered here.}, instead of the well studied case of $F(t)= F_0\big(\cos(\omega t)\big)$. From the form of Eq. \ref{eq1}, it can be seen that such a drive represents a periodically modulated non-Hermitian potential. Such potentials are realized by modulating the gain and loss of the system, which is achievable in quantum optical and superconducting-circuit systems due to the natural ability of photons to be excited and lost from such systems.

Inserting the PT-symmetric form of $F(t)$ into Eq. \ref{inteq}, and noting that for observable $\hat{A}$, $\hat{A}(t)=\hat{U}^\dagger(t)\hat{A}\hat{U}(t)$, we can explicitly write the dynamical evolution of the expectation values of the field quadratures for an initial coherent state $\ket{n_0}$. After some rearranging, these take the form

\begin{equation} \label{quads}
\begin{split}
\langle\hat{x}(t)\rangle&= \frac{\sqrt{2}}{\omega +1} \Big(\cos (t) \big(F_0+\omega n_0  +n_0\big)-F_0 \cos ( \omega t  )\Big)\\
\langle\hat{p}(t)\rangle&= \frac{-\sqrt{2}}{\omega +1} \Big(\sin (t) \big(F_0+\omega n_0  +n_0\big)+F_0 \sin ( \omega t  )\Big).
\end{split}
\end{equation}
A similar analysis can be used to show that the variance of the quadratures of the coherent initial state remain unchanged in time, and as a result, the quantum dynamics of the system are defined completely by the Wigner-space trajectories defined by Eqs. \ref{quads}.

{ The set of Wigner-space trajectories given by Eqs. \ref{quads} happen to represent a well known class of geometric curves known as hypotrochoids, which are in turn exactly equivalent to the set of real-space trajectories of all possible Foucault pendula. Geometrically, the hypotrochoid is generated by fixing a point a distance $d$ from the center of a circle of radius $r$, and tracing the path of that point as that circle is rolled around the inside of a larger circle of radius $R$. Here, the dual hypotrochoid can be defined parametrically in Euclidean coordinates $\{x,y\}$ by Eq. \ref{quads} 

\begin{figure}[t!]
\centering
\includegraphics[width=0.5\textwidth]{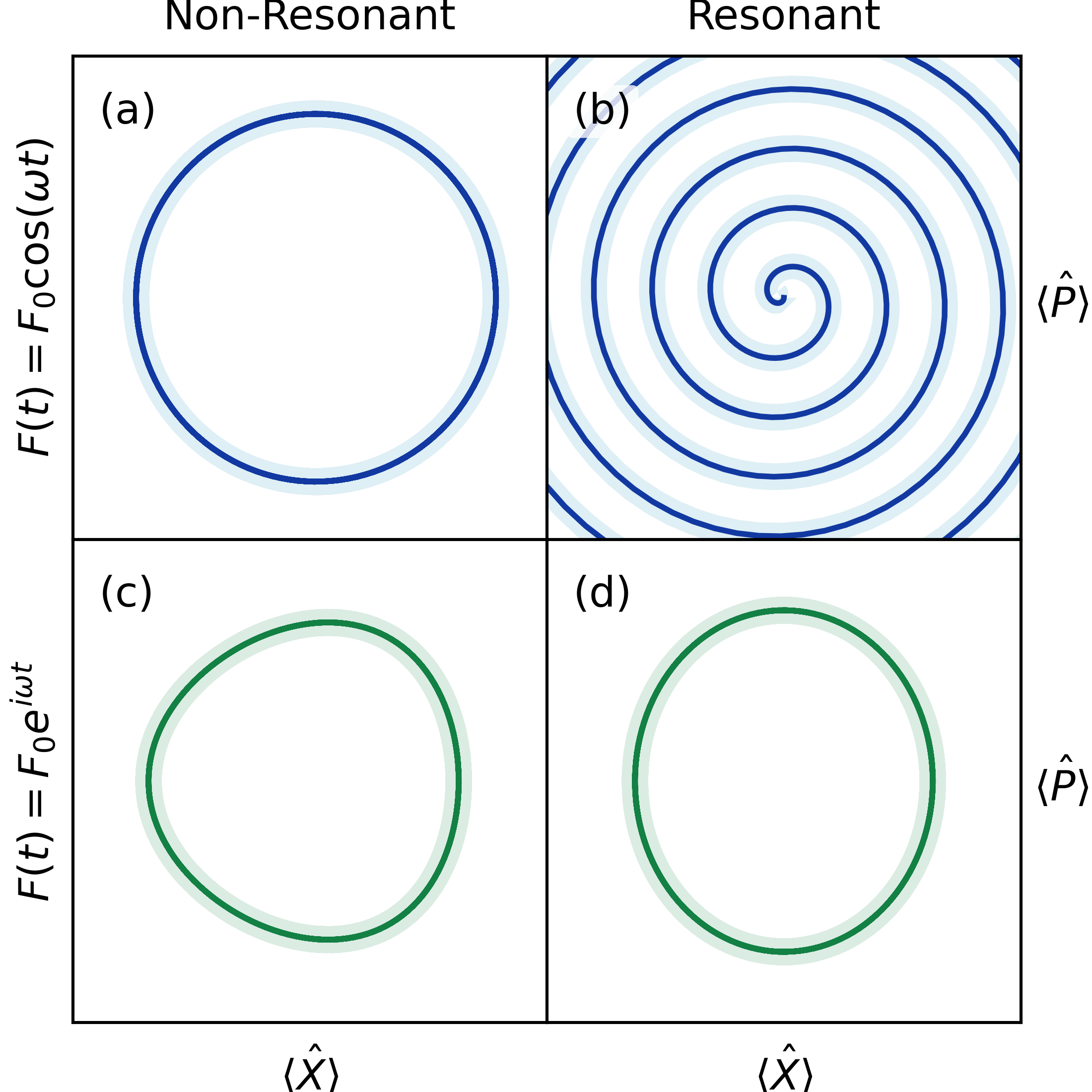}
\caption{(a)-(b): Quantum trajectories of a coherent state evolving under real driving of the form $F(t)=F_0\cos(\omega t)$. For suitably off-resonant driving, the trajectory is a closed ellipse (a), while near resonance the trajectory is unbounded, representing instability (b). For non-Hermitian, PT-symmetric driving, the trajectories are closed both on (d) and off resonance (c), representing the dynamical non-Hermitian stabilization of the resonantly driven quantum harmonic oscillator. For each plot, $F_0=1$. The resonant condition is $\omega=\omega_0$, while for the nonresonant case we show $\omega=2\omega_0$. The initial number states used in the plots are $10$, $0$, $8$, and $8$, respectively. However, qualitative results do not depend on particular parameters.}\label{curves}
\end{figure}

by replacing Wigner-space coordinates with real-space coordinates as 
$\{\langle\hat{x}(t)\rangle\rightarrow x,\langle\hat{p}(t)\rangle\rightarrow y\}$, and by setting $r= n_0+(F_0+n_0)/\omega$, $R= (\omega+1)(F_0+n_0(\omega+1))/\omega$, and $d=F_0$. This exact map allows one to select the parameters of the driven quantum harmonic oscillator such that the Wigner-space dynamics trace a particular subtype of the hypotrochoid, which include the hypocycloids, cardiods, rose curves, and the rounded polygons of any order. A few examples are shown in Fig. \ref{fig1}.}

Beyond representing an exact correspondence between seemingly unrelated quantum and classical oscillator phenomena, understanding the geometry of trajectory solutions allows us to better understand the type of phenomena that can be exhibited by the PT-driven quantum oscillator. While Fig. \ref{fig1} shows a number of exotic trajectories that a quantum state can take in the Wigner space, one of the most important class of solutions is the simple case of the circular trajectory. While the real-driven quantum oscillator only leads to circular trajectories for null driving strength, here we can exploit the Foucault map to derive a much wider class of circular trajectories, with arbitrary frequency: taking the set of circular solutions to the Foucault pendulum and employing the our mapping, it follows that by setting the strength of the non-Hermitian drive to $A=n_0(\omega-1)$, there exists an infinite set of circular quantum trajectories

\begin{equation} \label{circs}
\begin{split}
\langle\hat{x}(t)\rangle&= \sqrt{2}n_0\cos (\omega t)\\
\langle\hat{p}(t)\rangle&= -\sqrt{2}n_0\sin (\omega t),
\end{split}
\end{equation}
which represent the circular Wigner space orbits with arbitrary radius and frequency. Interestingly, this includes not only integer multiples of the natural frequency, but also includes fractional multiples of $\omega_0$. A case of a cicular orbit with twice the natural frequency is shown in Fig. \ref{fig1}.(a). The existence of this class of constant-energy solutions, to a quantum system driven far from equilibrium, is highly nontrivial.

Under real forcing at or near resonance, it is well understood that the quantum harmonic oscillator is unstable to an unbounded growth in the number operator expectation value $\langle \hat n\rangle = \langle \hat{a}^{\dagger}\hat{a}\rangle$ (\ref{quads_res}.(b)), 
representing a fundamental example of unstable quantum dynamics. However, under the non-Hermitian drive, the dynamics differ remarkably.
Setting $\omega=\omega_0$, Eqs. \ref{quads} reduce to

\begin{equation} \label{quads_res}
\begin{split}
\langle\hat{x}(t)\rangle_{res}&= \sqrt{2}n_0\cos(\omega_0 t)\\
\langle\hat{p}(t)\rangle_{res}&= -\sqrt{2}(n_0+F_0)\sin(\omega_0 t),
\end{split}
\end{equation}
which traces an ellipse in the Wigner-space \ref{quads_res}.(d). 
Therefore, simply by adding an imaginary component to the periodic drive (replacing $F(t)=F_0\cos(\omega_0t)$ by $F(t)=F_0\big(\cos(\omega_0t)+i\sin(\omega_0t)\big)$), the resonantly driven quantum oscillator, a fundamental example of quantum instability, is driven into dynamical stability.

\begin{figure}[t]
\centering
\includegraphics[width=0.5\textwidth]{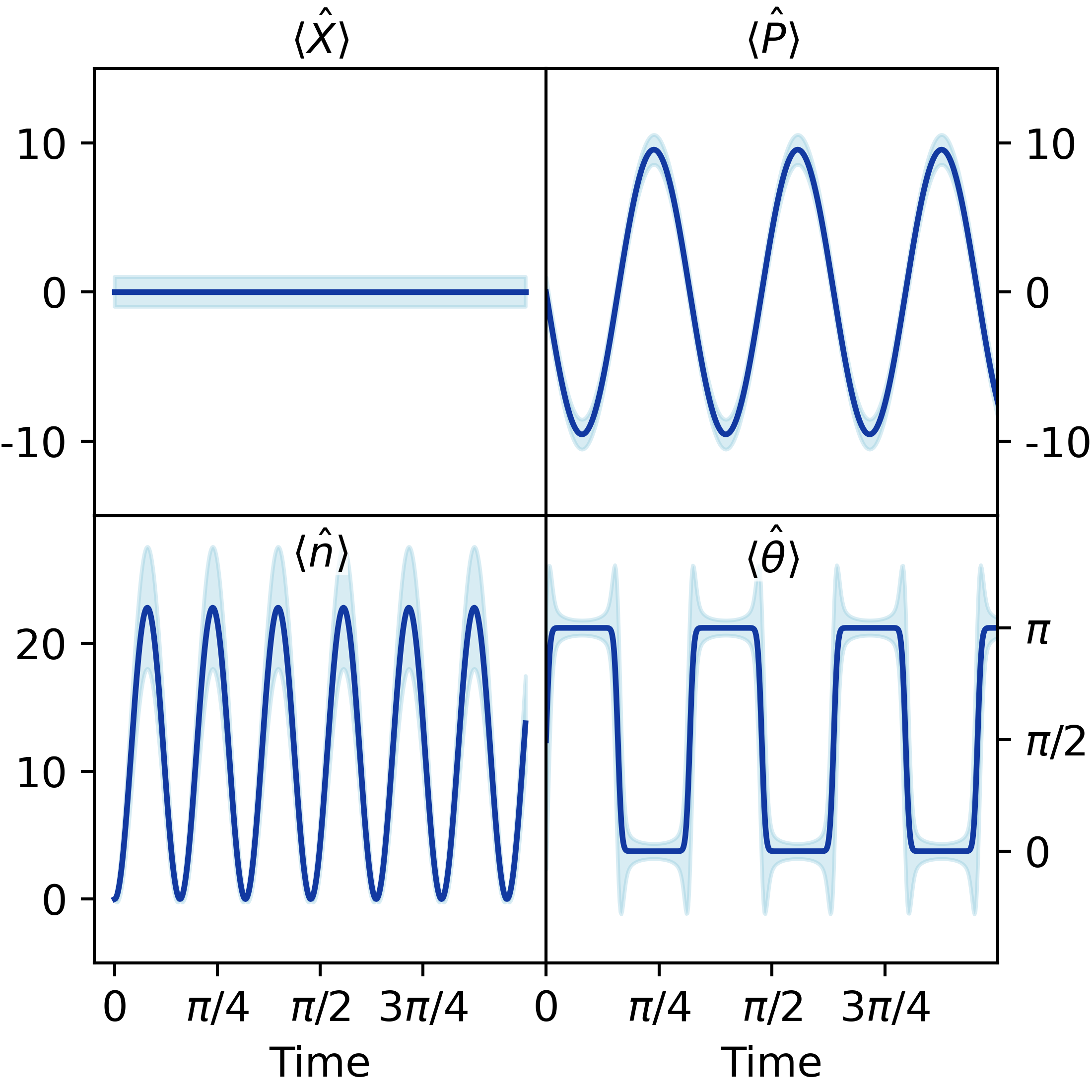}\label{stable}
\caption{Characteristic dynamics of the vacuum state driven at resonance. The expectation value (solid) of the position operator $\hat{x}(t)$ remains static, while that of the momentum operator $\hat{p}(t)$ evolves sinusoidally, exhibiting dynamics which are forbidden in classical physics. This evolution is characterized by regular pulses of occupation number, with pulses separated by temporal phase discontinuities.}
\end{figure}

Finally, we note that under this complex resonant driving, there is a particularly unusual set of modes that emerge when the initial state is a vacuum. Starting from the vacuum state $n_0=0$, the trajectories given by Eqs. \ref{quads_res} are reduced to
\begin{equation} \label{res0}
\begin{split}
\langle\hat{x}(t)\rangle_{res}&= 0\\
\langle\hat{p}(t)\rangle_{res}&= -\sqrt{2}F_0\sin(\omega_0 t).
\end{split}
\end{equation}
Such a trajectory has no analogue in classical phase space, as the momentum of the state oscillates periodically with amplitude set arbitrarily by $F_0$, while the position of the state remains at rest. In the dual Foucault system, this corresponds to the trivial case of one-dimensional motion (ie, a regular pendulum). Interestingly, the time-dependent expectation value of the Pegg-Barnett phase operator \cite{pegg1989phase} $\hat{\theta}$ of this mode can be found to take the form

\begin{equation}
    \langle \hat{\theta}(t) \rangle=\arctan\big( -F_0\sin(\omega_0 t )\big),
\end{equation}
so that as the drive strength $F_0$ increases, $\langle\hat{\theta}(t)\rangle$ asymptotically approaches a square wave, flipping discontinuously between $0$ and $\pi$. Formally, this square wave has the form
\begin{equation}
    \lim_{F_0\rightarrow\infty}\big(\langle\hat{\theta}(t)\rangle\big) = \frac{\pi}{2}  \bigg(\mathrm{Sgn}\big[\sin (\omega_0 t)\big]+1\bigg).
\end{equation}
Physically, this represents the excitation of a pulsed cavity mode, with pulses separated by topological phase defects resulting in a train of temporal phase solitons (ie, instantons) separated by timescales on the order of the optical frequency. Beyond inherent theoretical interest, such a quantum optical mode might, in close analogy with the flip-flop signals used in classical electronics, find use as a naturally discrete phase reference.

The author thanks Natalia Berloff and the University of Cambridge Centre for Mathematical Sciences, where this work began, for their generous hospitality. The author also thanks Eddy Timmermans for his encouragement, Layton Hall, and Katarzyna Krzyzanowska for useful conversations regarding experimental considerations, and Rebecca Alperin for the close reading and copyediting this text.  This work was supported by EPSRC grant EP/R014604/1, and by Los Alamos National
Laboratory LDRD program grant 20230865PRD3.

\end{document}